\documentclass[a4paper,11pt]{article}
\usepackage{pos}
\usepackage{subfigure}

\title{Search for new physics with charm rare decays at BESIII}

\author*[a]{Yonghua Zhan}
\onbehalf{on behalf of the BESIII Collaboration}

\affiliation[a]{School of Physics, Sun Yat-sen University,\\
Guangzhou 510275, China}

\emailAdd{zhanyh6@mail2.sysu.edu.cn}

\abstract{Search for new physics with charm rare decays at BESIII.
The BESIII experiment has collected 2.6 billion $\psi(3686)$ events, 10 billion $J/\psi$ events, $20 fb^{-1}$ D meson pairs at 3.773 GeV, and $7.33 fb^{-1}~D_{s}D_{s}^{*}$ events from 4.128 to 4.226 GeV. The huge data samples allow us to search for rare processes in charm hadron decays. In this paper, we report the FCNC decay in $J/\psi\to D^0\mu^+\mu^-$, $J/\psi\to D^0\gamma$ and $D_{s}^{+}\to h(h')e^+e^-$. The search for $J/\psi$ weak decays containing a D meson, $J/\psi\to D_{s}^{-}\pi^{+}$, and $J/\psi\to D_{s}^{-}\rho^{+}$ will also be presented.}

\FullConference{The European Physical Society Conference on High Energy Physics (EPS-HEP2025)\\
7-11 July 2025\\
Marseille, France\\}

\begin{document}
\maketitle

\section{Introduction}
Compared with processes where the Standard Model (SM) contribution is dominant, processes with highly suppressed or forbidden SM contributions can provide better sensitivity to new physics. This paper focuses on processes with highly suppressed SM contributions, including flavor-changing neutral current~(FCNC) processes and $J/\psi$ weak decays. The BESIII experiment, with its high statistics, clean background, and good resolution, is suitable for searching for new physics in these channels.

In this paper, we introduce the searches for new physics with charm rare decays at BESIII. In Section~\ref{sec:detector}, we describe the BESIII detector and the data samples used in this analysis. The searches for FCNC decay $J/\psi\to D^{0}\mu^{+}\mu^{-}$~\cite{BESIII:2025wsu}, $J/\psi\to \gamma D^{0}$~\cite{BESIII:2024kkx}, and $D_{s}^{+}\to h(h')e^{+}e^{-}$~\cite{BESIII:2024nrw} will be introduced in Section~\ref{sec:FCNC}. The search for J/psi weak decays containing a D meson~\cite{BESIII:2023qpx}, $J/\psi\to D_{s}^{-}\pi^{+}$, and $J/\psi\to D_{s}^{-}\rho^{+}$~\cite{BESIII:2025rjn} will also be presented in Section~\ref{sec:jpsi}. Finally, a summary of new physics with charm search at BESIII will be presented in Section~\ref{sec:summary}.

\section{BESIII Detector and data samples}
\label{sec:detector}
The BESIII detector~\cite{BESIII:2009fln} records symmetric $e^+e^-$ collisions 
provided by the BEPCII storage ring~\cite{Yu:IPAC2016-TUYA01} in the center-of-mass energy range from 1.84 to 4.95~GeV, with a peak luminosity of $1.1 \times 10^{33}\;\text{cm}^{-2}\text{s}^{-1}$ achieved at $\sqrt{s} = 3.773\;\text{GeV}$. 
BESIII has collected large data samples in this energy region~\cite{Ablikim:2019hff, EcmsMea, EventFilter, Liao:2025lth}. The cylindrical core of the BESIII detector covers 93\% of the full solid angle and consists of a helium-based multilayer drift chamber~(MDC), a time-of-flight system~(TOF), and a CsI(Tl) electromagnetic calorimeter~(EMC),
which are all enclosed in a superconducting solenoidal magnet providing a 1.0~T magnetic field.
The magnetic field was 0.9~T in 2012, which affects 11\% of the total $J/\psi$ data.
The solenoid is supported by an octagonal flux-return yoke with resistive plate counter muon identification modules interleaved with steel. 
The charged-particle momentum resolution at $1~{\rm GeV}/c$ is $0.5\%$, and the 
${\rm d}E/{\rm d}x$ resolution is $6\%$ for electrons from Bhabha scattering. The EMC measures photon energies with a resolution of $2.5\%$ ($5\%$) at $1$~GeV in the barrel (end cap) region. The time resolution in the plastic scintillator TOF barrel region is 68~ps, while that in the end cap region was 110~ps. The end cap TOF
system was upgraded in 2015 using multigap resistive plate chamber technology, providing a time resolution of 60~ps, which benefits 87\% of the data used in this analysis~\cite{etof}.

Monte Carlo~(MC) simulated data samples produced with a {\sc geant4}-based~\cite{geant4} software package, which includes the geometric description of the BESIII detector~\cite{bes:unity, geo1, geo2, Li:2024pox, Song:2025pnt} and the detector response, are used to determine detection efficiencies and to estimate backgrounds. The simulation models the beam energy spread and initial state radiation~(ISR) in the $e^+e^-$ annihilations with the generator {\sc kkmc}~\cite{ref:kkmc, ref:kkmc2}. 
All particle decays are modeled with {\sc evtgen}~\cite{ref:evtgen, ref:evtgen2} using BFs either taken from the Particle Data Group~\cite{pdg:2024}, when available, or otherwise estimated with {\sc lundcharm}~\cite{ref:lundcharm, ref:lundcharm2}. Final state radiation~(FSR) from charged final state particles is incorporated using the {\sc photos} package~\cite{photos2}.

\section{Searches for FCNC processes}
\label{sec:FCNC}
FCNC transitions such as $c\to u$ manifest solely through loop-level interactions due to the Glashow-Iliopoulos-Maiani~(GIM) mechanism~\cite{Glashow:1970gm}, which result in low branching fractions (BFs) of the corresponding decays. The FCNC decay $J/\psi\to D^{0}l^{+}l^{-}$ is expected to have a BF on the order of $10^{-13}$ in the SM~\cite{Wang:2008pq}. In comparison, the FCNC process $J/\psi\to D^{0}\gamma$ is expected with a 1 or 2 orders of magnitude larger BF, due to the presence of one fewer decay vertex. Nevertheless, various new physics (NP) models beyond the SM, such as the Top-Color model~\cite{hill:1995}, the supersymmetric extensions of the SM with or without R-parity violation~\cite{Aulakh:1982yn}, and the two-Higgs doublet model~\cite{Glashow:1976nt}, suggest these BFs may be enhanced to exceed the SM prediction. Any enhancement of the BF with respect to the SM would be a strong indication of NP.

\subsection{Searches for $J/\psi\to D^{0}\mu\mu$ and $J/\psi\to D^{0}\gamma$}
\label{sec:etau}
\begin{figure*}[htbp]
\centering
\subfigure[$J/\psi\to D^{0}\mu\mu$]{\includegraphics[width=0.3\textwidth]{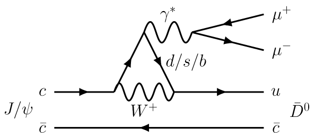}}
\subfigure[$J/\psi\to D^{0}\gamma$]{\includegraphics[width=0.3\textwidth]{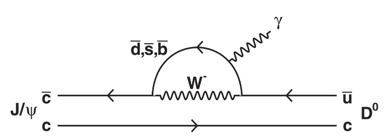}}
\caption{Feynman diagrams in the SM.}
\label{fig:fm1}
\end{figure*}

\begin{figure*}[htbp]
\centering
\subfigure[]{\includegraphics[width=0.22\textwidth]{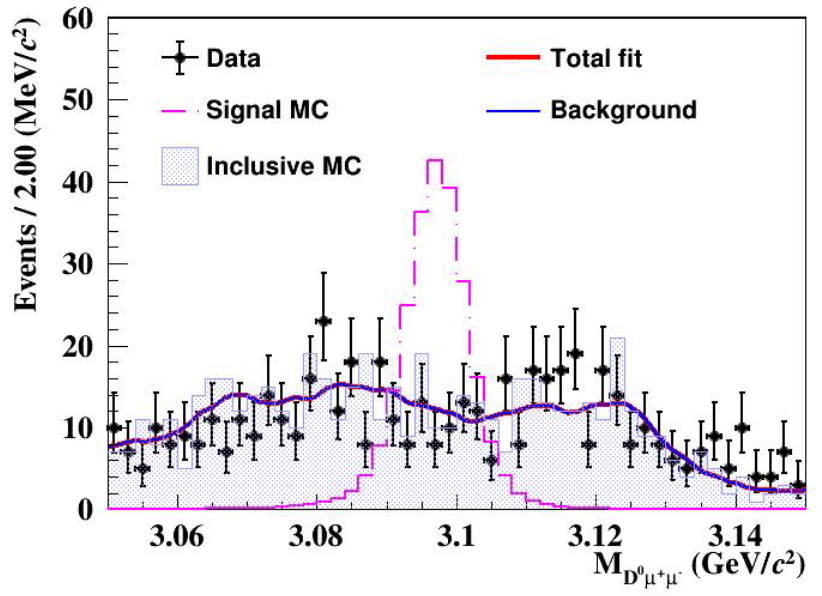}}
\subfigure[]{\includegraphics[width=0.22\textwidth]{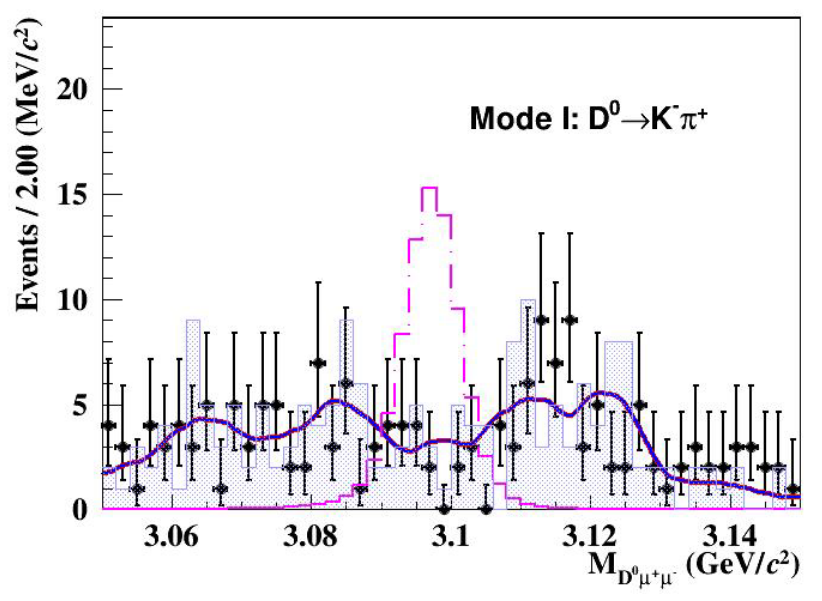}}
\subfigure[]{\includegraphics[width=0.22\textwidth]{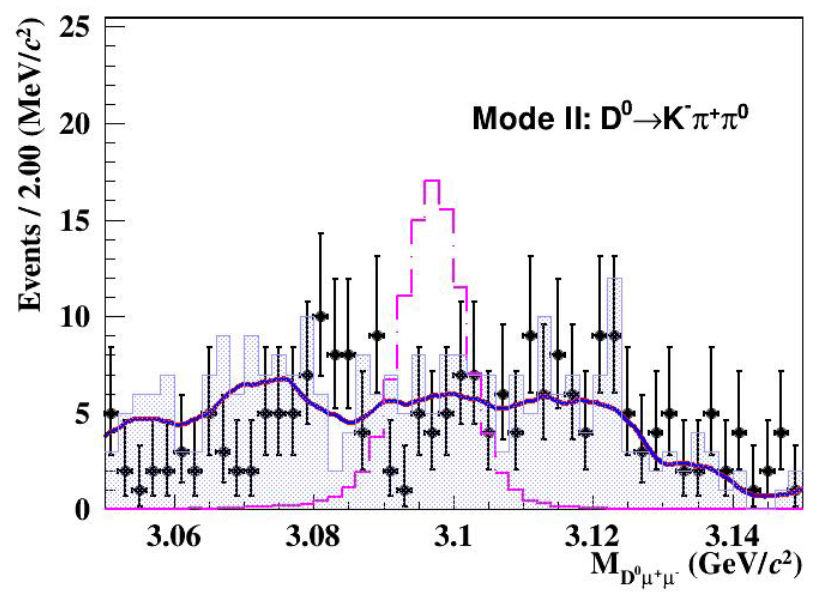}}
\subfigure[]{\includegraphics[width=0.22\textwidth]{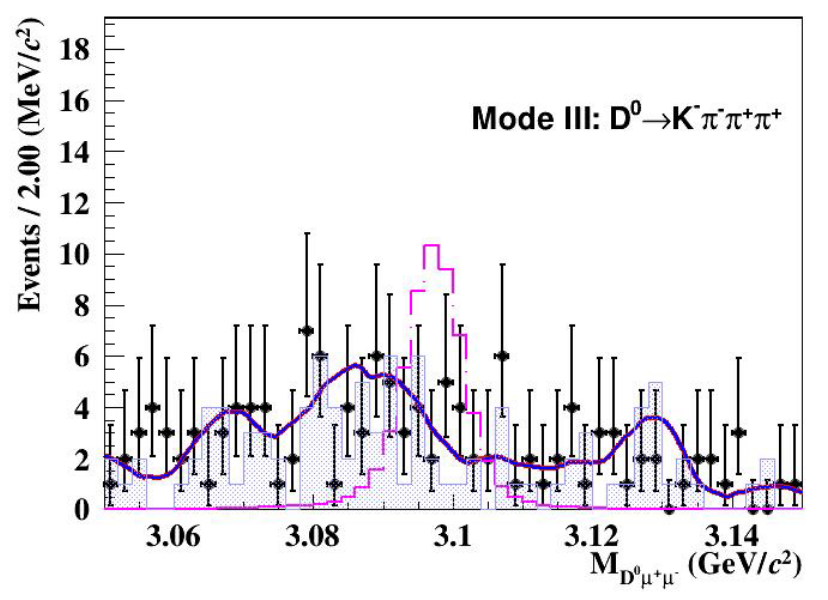}}
\caption{The distributions of $M_{D^{0}\mu^{+}\mu^{-}}$ for $J/\psi\to D^{0}\mu^{+}\mu^{-}$ of the selected candidates in data, signal MC sample, and inclusive MC sample. The simultaneous fit result is shown in the top left, and the individual fit result for each mode is shown in the other three sub-figures, respectively. The black dots with error bars are data, the magenta dotted-dashed line shows the shape of the signal MC sample scaled to $\mathcal{B}(J/\psi\to D^{0}\mu^{+}\mu^{-}) = 1.0 \times 10^{-6}$ for the three tags. The blue shaded histogram is the inclusive MC sample, the red line is the fit result, and the blue solid lines are the fitted background.}
\label{fig:Dmumu1}
\end{figure*}

\begin{figure*}[htbp]
\centering
\subfigure[]{\includegraphics[width=0.25\textwidth]{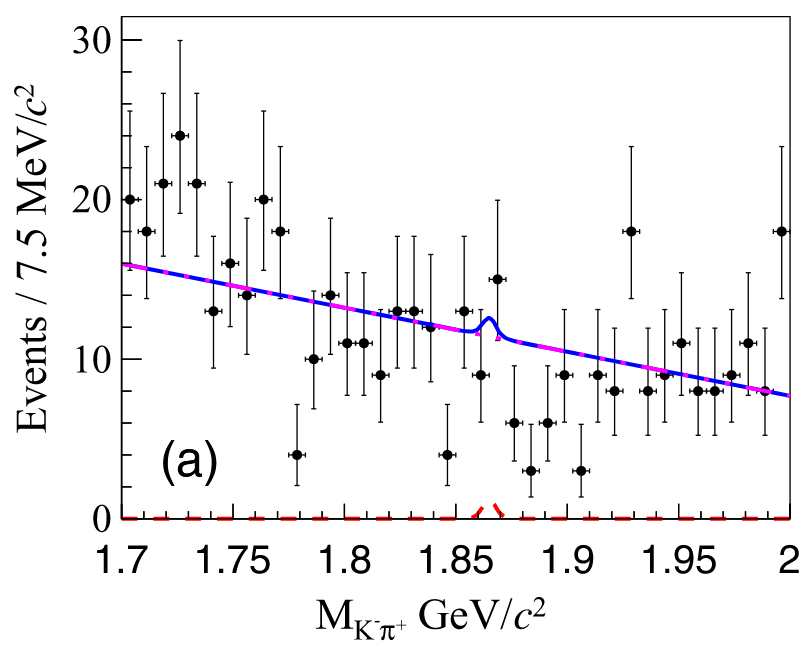}}
\subfigure[]{\includegraphics[width=0.25\textwidth]{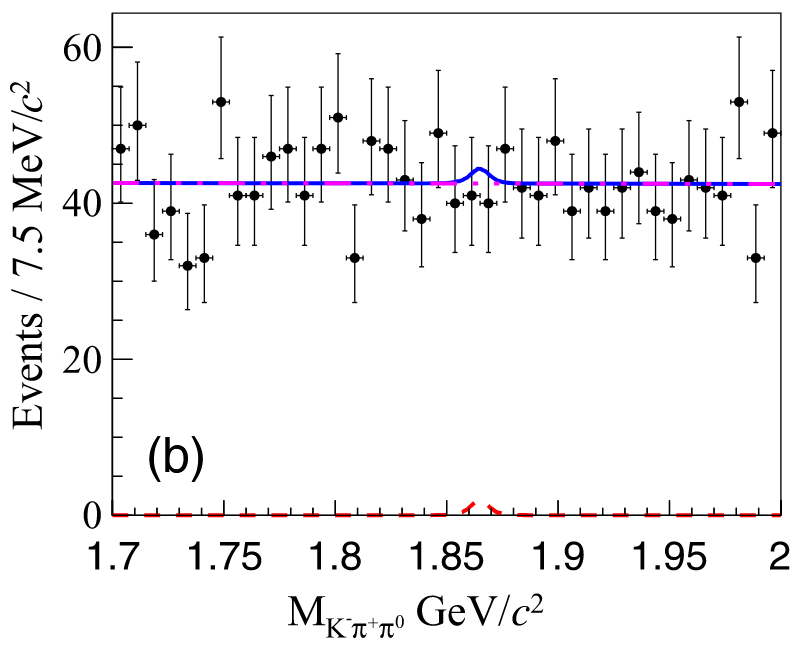}}
\subfigure[]{\includegraphics[width=0.25\textwidth]{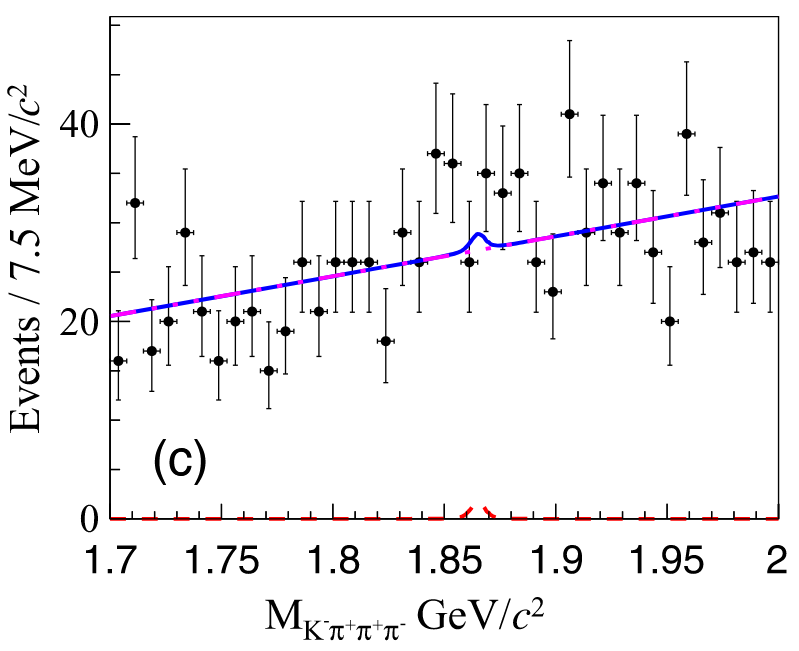}}
\caption{The unbinned simultaneous fit on the invariant mass distributions for $J/\psi\to D^{0}\gamma$ of modes (a) I, (b) II, and (c) III, where the dots with error bar are data, the blue lines are the fit result, and the red dotted and purple curves are the fitted signal and background, respectively.}
\label{fig:Dgamma1}
\end{figure*}

The searches for FCNC decay $J/\psi\to D^{0}\mu\mu$ and $J/\psi\to D^{0}\gamma$ are performed with the $(10087\pm44)\times10^{6}$ $J/\psi$ data events~\cite{bes3:totJpsiNumber}. These decays are typical FCNC processes that can occur through possible Feynman diagrams such as those shown in Figure~\ref{fig:fm1}. Three tag modes are used to reconstruct $D^{0}$ meson: $D^{0}\to K^{-}\pi^{+}$ (Mode I), $D^{0}\to K^{-}\pi^{+}\pi^{0}$ with $\pi^{0}\to\gamma\gamma$ (Mode II), and $D^{0}\to K^{-}\pi^{-}\pi^{+}\pi^{+}$ (Mode III), which have large BFs.
To extract the BFs, an unbinned simultaneous maximum-likelihood fit is carried out to the selected samples of the three $D^{0}$ decay modes by sharing the same decay BF of $J/\psi\to D^{0}\mu\mu$~($J/\psi\to D^{0}\gamma$), as shown in Figure~\ref{fig:Dmumu1} and Figure~\ref{fig:Dgamma1}.

For $J/\psi\to D^{0}\mu^{+}\mu^{-}$, no significant signal is observed, and the upper limits on BFs are set to be $\mathcal{B}
(J/\psi\to D^{0}\mu^{+}\mu^{-})= 1.1 \times 10^{-7}$  at the 90\% C.L. after considering all systematic uncertainties. This is the first search for a charmonium FCNC process with muons in the final state. This result is compatible with the SM prediction on its BF of $10^{-13}$.

For $J/\psi\to D^{0}\gamma$, no significant signal is observed, and the upper limits on BFs are set to be $\mathcal{B}(J/\psi\to D^{0}\gamma) = 9.1 \times 10^{-8}$ at the 90\% C.L. after considering all systematic uncertainties. Although our measurement does not yet reach the precision predicted by the SM, it provides a valuable reference for probing various new-physics models and restricting their parameter space.

\subsection{Search for Rare Decays $D_{s}^{+}\to h(h')e^{+}e^{-}$}
\label{sec:Ds}
In the SM, the rare decays $D_{s}^{+}\to h(h')e^{+}e^{-}~(h=\pi,K)$, involve both short-distance~(SD) and long-distance~(LD) contributions. The SD contribution proceeds via the FCNC transition $c\to u\ell^{+}\ell^{-}$. It is loop-induced and strongly suppressed by the GIM mechanism, which is more effective in the charm sector than in the strange or bottom sectors. Consequently, SD-only branching fractions are as low as $10^{-9}$. The LD contributions, occurring via a radiated photon or an intermediate meson decaying to dileptons, dominate the decays of $D_{s}^{+}\to h(h')e^{+}e^{-}$ and raise the BFs to the order of $10^{-6}$.

We measure the LD contributions of $D_{s}^{+}\to\pi^{+}\phi$ with $\phi\to e^{+}e^{-}$, $D_{s}^{+}\to\rho^{+}\phi$ with $\rho^{+}\to\pi^{+}\pi^{0}$ and  $\phi\to e^{+}e^{-}$, and search for the four-body decays of $D_{s}^{+}\to \pi^{+}\pi^{0}e^{+}e^{-}$, $D_{s}^{+}\to K^{+}\pi^{0}e^{+}e^{-}$, $D_{s}^{+}\to K_{S}^{0}\pi^{+}e^{+}e^{-}$ using data samples corresponding to an integrated luminosity of $7.33 fb^{-1}$ accumulated with the BESIII detector at $e^{+}e^{-}$ center-of-mass (c.m.) energies in the range $\sqrt{s}=4.128-4.226\mathrm{\ Ge\kern -0.1em V}$. The $D_{s}^{+}$ candidates are selected from the process of $e^{+}e^{-}\to D_{s}^{*\pm}(\to D_{s}\gamma)D_{s}^{\mp}$ in this work.

\begin{figure*}[htbp]
\centering
\subfigure[]{\includegraphics[width=0.3\textwidth]{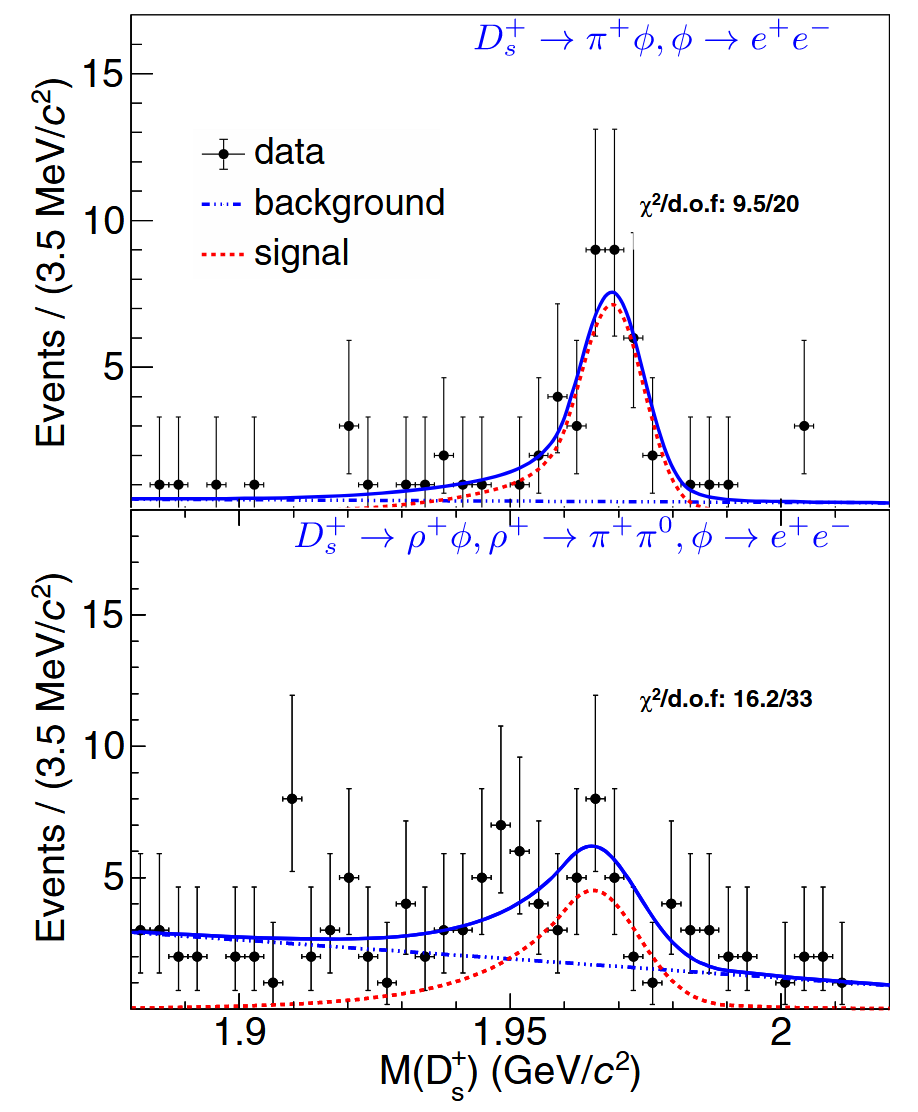}}
\subfigure[]{\includegraphics[width=0.3\textwidth]{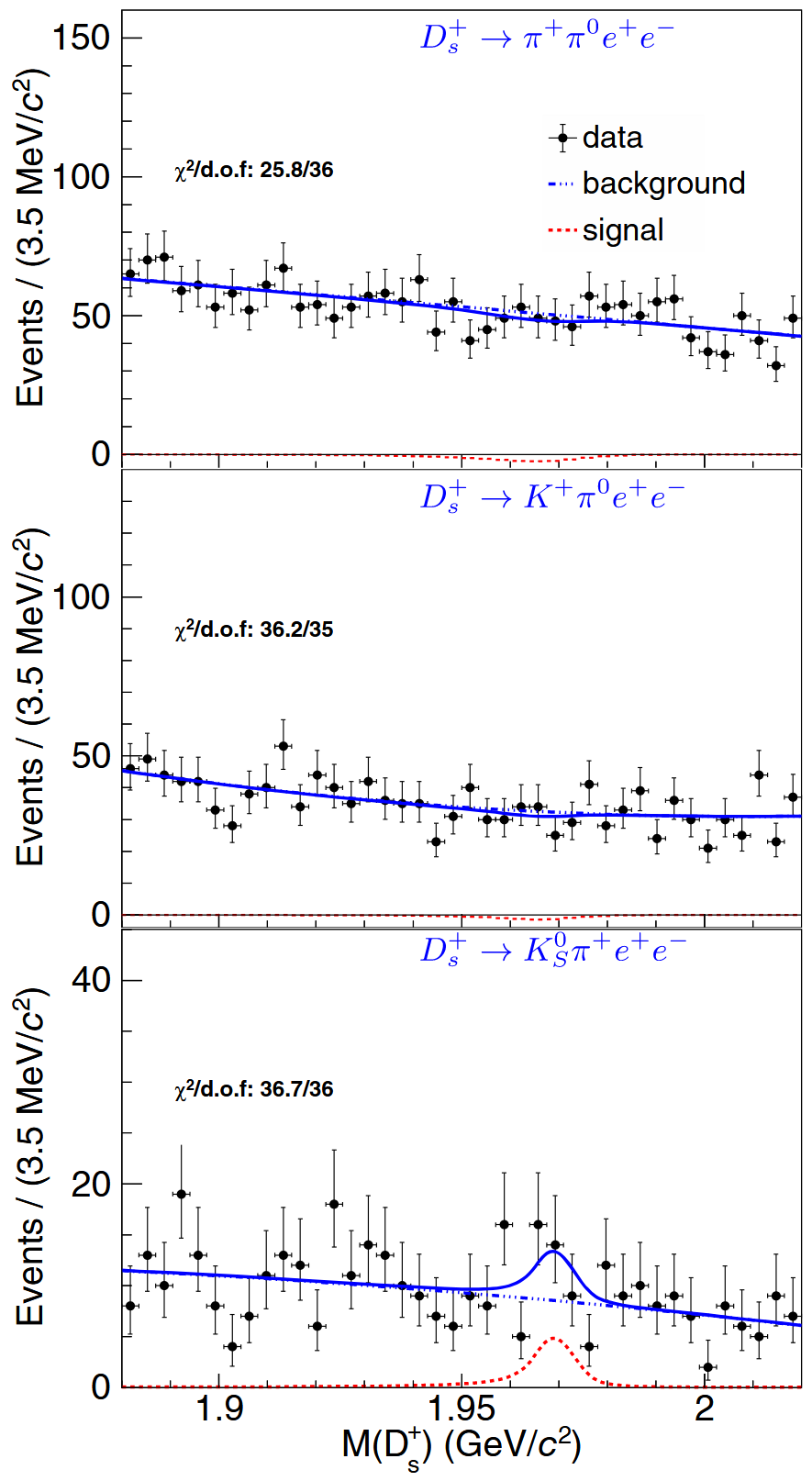}}
\caption{Fits to the $M(D_{s}^{+})$ distributions for (a) $D_{s}^{+}\to\pi^{+}\phi(\to e^{+}e^{-})$, $D_{s}^{+}\to\rho^{+}(\to\pi^{+}\pi^{0})\phi(\to e^{+}e^{-})$, (b) $D_{s}^{+}\to \pi^{+}\pi^{0}e^{+}e^{-}$, $D_{s}^{+}\to K^{+}\pi^{0}e^{+}e^{-}$, and $D_{s}^{+}\to K_{S}^{0}\pi^{+}e^{+}e^{-}$. The signals are shown as the magenta dashed curves. The blue long-dashed curves are the combinatorial background components, and the dots with error bars are data.}
\label{fig:Ds2}
\end{figure*}

The signal yield for each $D_{s}^{+}$ decay mode is extracted independently by performing an unbinned maximum-likelihood fit to the corresponding invariant-mass spectrum in the range $1.88$–$2.02\,\text{GeV}/c^{2}$. Figure~\ref{fig:Ds2} shows the data distributions and fit results for these decay modes. The $D_{s}^{+}\to\pi^{+}\phi$, $\phi\to e^{+}e^{-}$ decay is observed with a statistical significance of $7.8\sigma$. Evidence of the $D_{s}^{+}\to\rho^{+}\phi$, $\phi\to e^{+}e^{-}$ decay is found for the first time with a statistical significance of $4.4\sigma$. The BFs of these two decays are measured to be $\mathcal{B}(D_{s}^{+}\to\pi^{+}\phi, \phi\to e^{+}e^{-})=(1.17_{-0.21}^{+0.23}\pm0.03)\times10^{-5}$ and $\mathcal{B}(D_{s}^{+}\to\rho^{+}\phi, \phi\to e^{+}e^{-})=(2.44_{-0.62}^{+0.67}\pm0.16)\times10^{-5}$, where the first uncertainties are statistical and the second systematic. No significant signal of the four-body rare decays is observed, and the upper limits on the BFs of these decays are set to be $\mathcal{B}(D_{s}^{+}\to \pi^{+}\pi^{0}e^{+}e^{-})<7.0\times10^{-5}$, $\mathcal{B}(D_{s}^{+}\to K^{+}\pi^{0}e^{+}e^{-})<7.1\times10^{-5}$, and $\mathcal{B}(D_{s}^{+}\to K_{S}^{0}\pi^{+}e^{+}e^{-})<8.1\times10^{-5}$, at the 90\% confidence level. These results represent the first upper limits on the BFs of these decays.

\section{Searches for $J/\psi$ weak decays}
\label{sec:jpsi}
Via the weak interaction, the $J/\psi$ can potentially decay into a single charm meson via such as D, accompanied by some non-charm mesons. Some of them have been searched for by BESIII\cite{Li:2024moj, bes:2008, bes3:2014xbo, bes3:2014, bes3:2017, bes3:2021, BESIII:2023fqz, BESIII:2022ibp}. The inclusive BF of charmonium rare weak decays is predicted to be of the order of $10^{-8}$ in the SM~\cite{verma:1990, Sanchis:1994, Sanchis:1993, sharma:1999, wang:2008a, wang:2008b, shen:2008, dhir:2013, ivanov:2015, tian:2017, Sun:2023uyn, Meng:2024nyo}. If a signal for one of these decays is observed with BFs in the range of $10^{-8}$ to $10^{-6}$, it would indicate new physics beyond the SM.

\begin{figure*}[htbp]
\centering
\subfigure[$J/\psi\to \bar{D}^{0}\pi^{0}$, $J/\psi\to \bar{D}^{0}\eta$, and $J/\psi\to \bar{D}^{0}\rho^{0}$]{\includegraphics[width=0.22\textwidth]{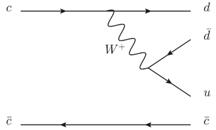}}
\subfigure[$J/\psi\to D^{-}\pi^{+}$ and $J/\psi\to D^{-}\rho^{+}$]{\includegraphics[width=0.22\textwidth]{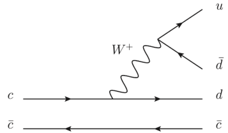}}
\subfigure[$J/\psi\to D_{s}^{-}\pi^{+}$ and $J/\psi\to D_{s}^{-}\rho^{+}$]{\includegraphics[width=0.28\textwidth]{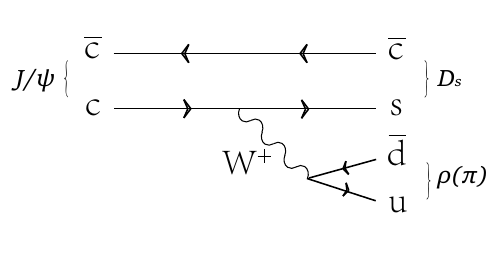}}
\caption{Feynman diagrams in the SM.}
\label{fig:fm2}
\end{figure*}

Using a sample of $(10087\pm44)\times10^{6}$ $J/\psi$ events collected at the BESIII detector, we search for the weak decays $J/\psi\to \bar{D}^{0}\pi^{0}$, $J/\psi\to \bar{D}^{0}\eta$, $J/\psi\to \bar{D}^{0}\rho^{0}$, $J/\psi\to D^{-}\pi^{+}$, $J/\psi\to D^{-}\rho^{+}$, $J/\psi\to D_{s}^{-}\pi^{+}$, and $J/\psi\to D_{s}^{-}\rho^{+}$. Figure~\ref{fig:fm2} shows the Feynman diagrams for these decay modes in the SM.
\subsection{Search for $J/\psi$ weak decays containing a D meson}
\label{sec:Dmeson}
\begin{figure*}[htbp]
\centering
\subfigure[]{\includegraphics[width=0.25\textwidth]{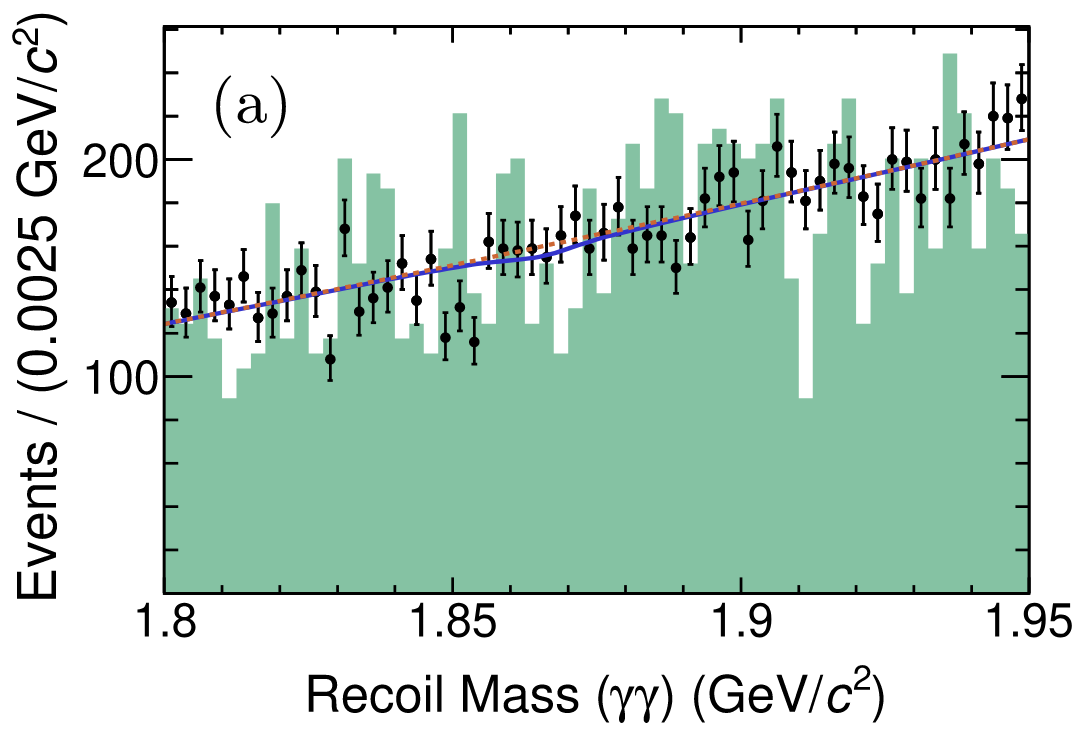}}
\subfigure[]{\includegraphics[width=0.25\textwidth]{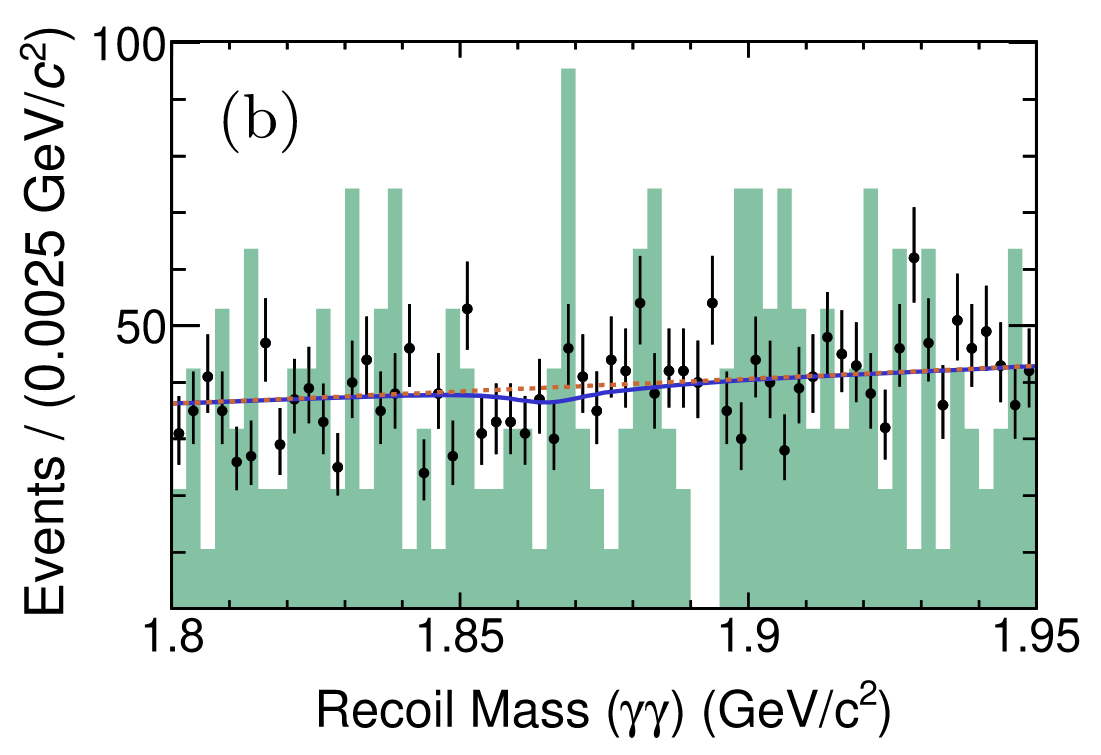}}
\subfigure[]{\includegraphics[width=0.25\textwidth]{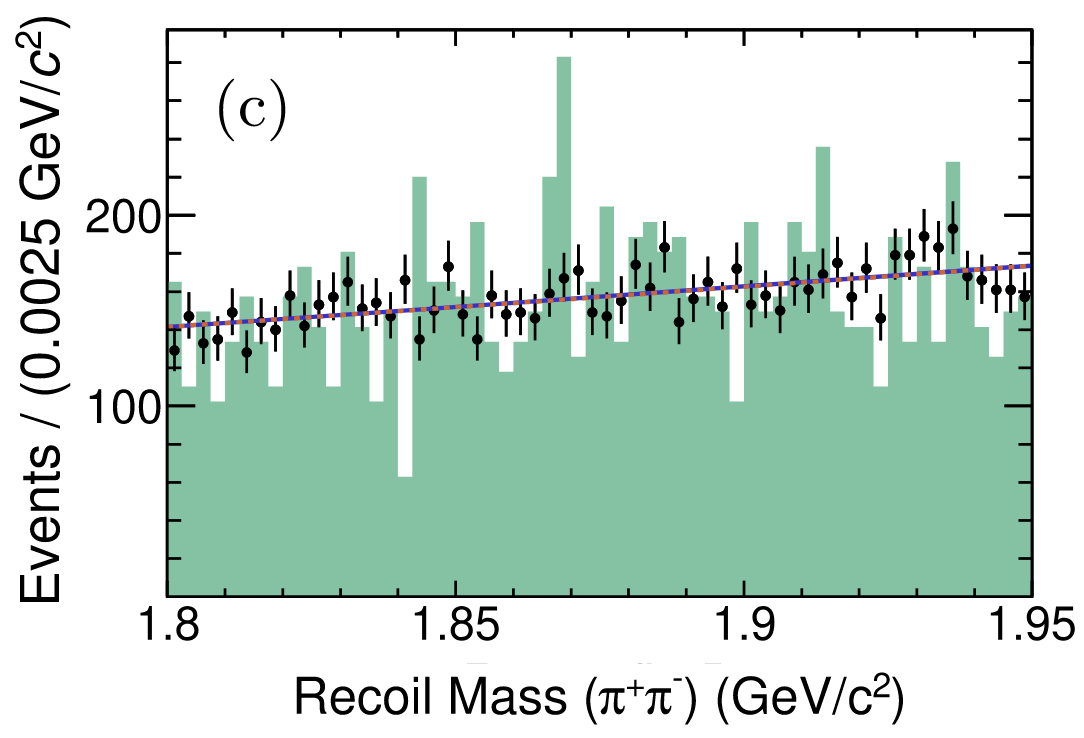}}
\subfigure[]{\includegraphics[width=0.25\textwidth]{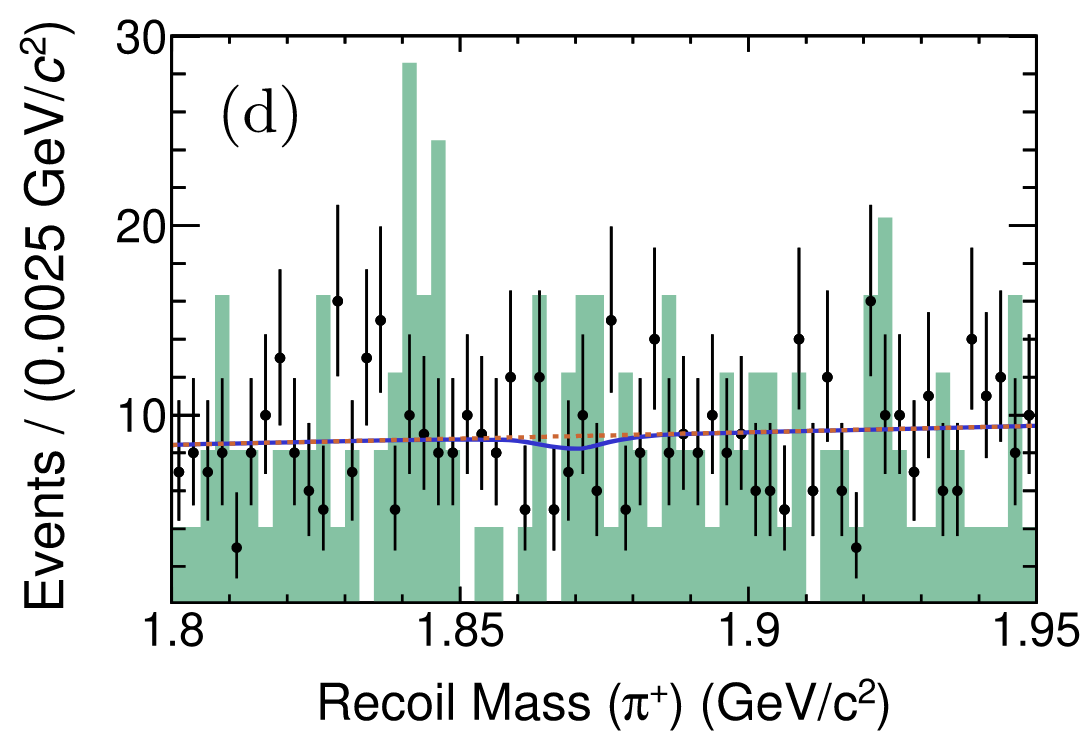}}
\subfigure[]{\includegraphics[width=0.25\textwidth]{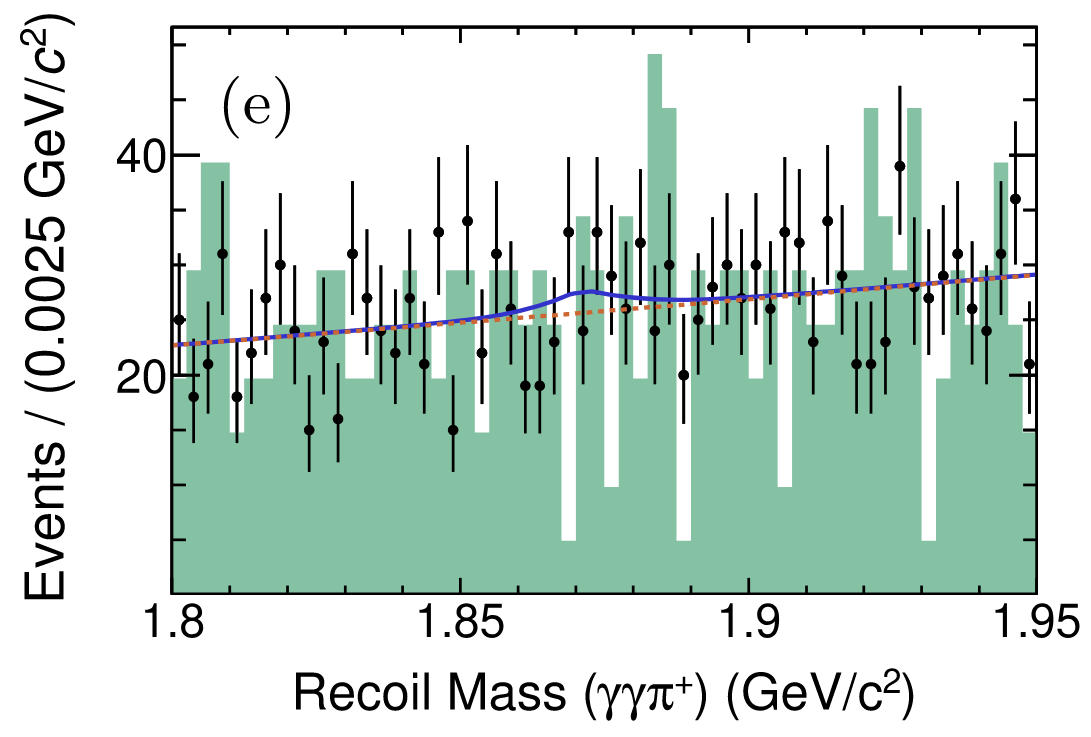}}
\caption{Fits of the accepted candidates to the recoiling mass spectra for (a) $J/\psi\to \bar{D}^{0}\pi^{0}$, (b) $J/\psi\to \bar{D}^{0}\eta$, (c) $J/\psi\to \bar{D}^{0}\rho^{0}$, (d) $J/\psi\to D^{-}\pi^{+}$, and (e) $J/\psi\to D^{-}\rho^{+}$. The dots with error bars are data, and the orange dotted lines are polynomial functions describing the background. The blue solid curves are the total fits. The inclusive MC samples are shown by the green filled histograms.}
\label{fig:D}
\end{figure*}

To avoid high background from conventional $J/\psi$ hadronic decays, we tag $\bar D^{0}$ and $D^{-}$ mesons through the semileptonic channels $\bar D^{0}\to K^{+}e^{-}\bar\nu$ and $D^{-}\to K_{S}^{0}e^{-}\bar\nu$ ($K_{S}^{0}\to\pi^{+}\pi^{-}$). Since the neutrino is undetectable at BESIII, the $\bar{D}^{0}$ and $D^{-}$ mesons cannot be fully reconstructed from their decay products. However, the $\bar{D}^{0}$ and $D^{-}$ mesons can be identified via the mass recoiling against the accompanying light hadron $\pi^{0}$, $\eta$, $\rho^{0}$, $\pi^{+}$, and $\rho^{+}$ with $\pi^{0}/\eta\to\gamma\gamma$, $\rho^{0}\to\pi^{+}\pi^{-}$, and $\rho^{+}\to\pi^{+}\pi^{0}$, respectively. Figure~\ref{fig:D} shows the recoiling mass spectra of the accepted candidates for these decays. No significant signal is observed in any of the decay modes. The upper limits at the 90\% CL on the branching fractions are determined to be: $\mathcal{B}(J/\psi\to \bar{D}^{0}\pi^{0}+c.c.)<4.7\times10^{-7}$, $\mathcal{B}(J/\psi\to \bar{D}^{0}\eta+c.c.)<6.8\times10^{-7}$, $\mathcal{B}(J/\psi\to \bar{D}^{0}\rho^{0}+c.c.)<5.2\times10^{-7}$, $\mathcal{B}(J/\psi\to D^{-}\pi^{+}+c.c.)<7.0\times10^{-8}$, and $\mathcal{B}(J/\psi\to D^{-}\rho^{+}+c.c.)<6.0\times10^{-7}$.

\subsection{Search for $J/\psi$ weak decays }
\label{sec:Dsrhopi}
\begin{figure*}[htbp]
\centering
\subfigure[]{\includegraphics[width=0.3\textwidth]{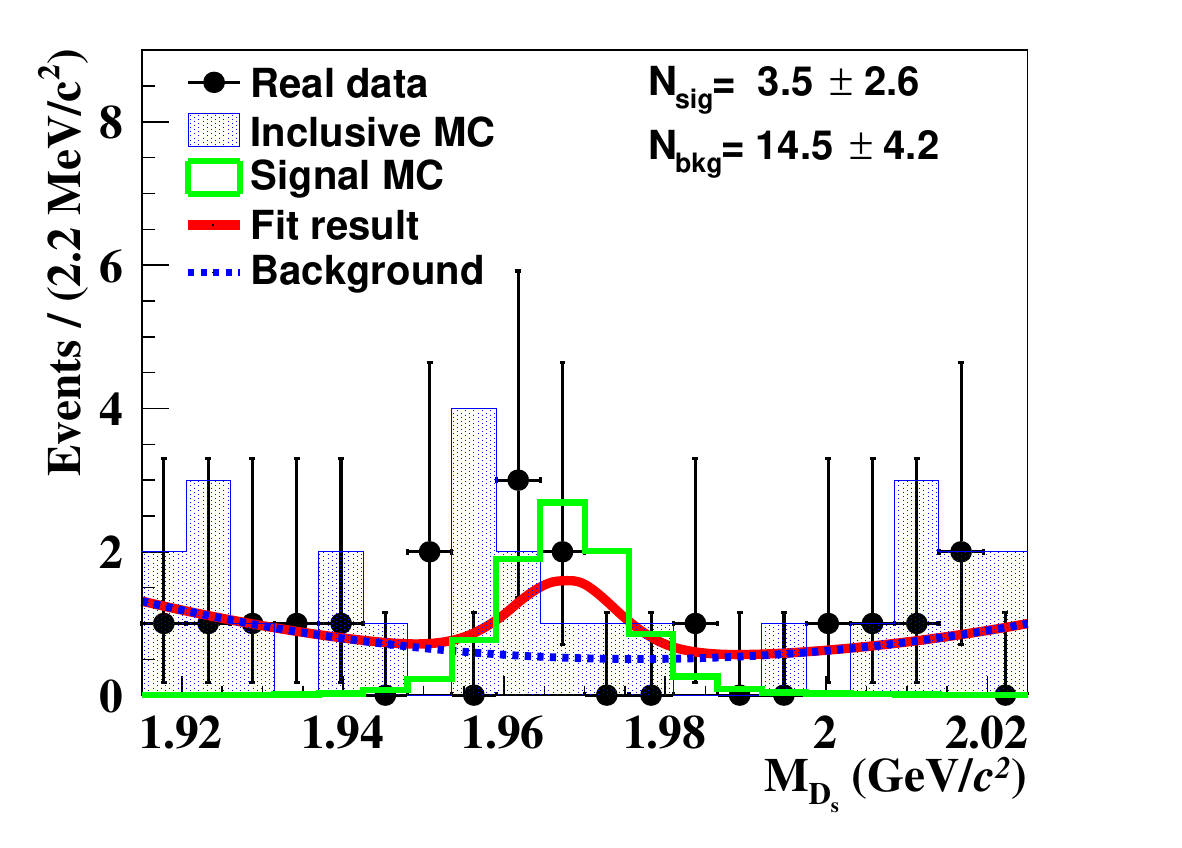}}
\subfigure[]{\includegraphics[width=0.3\textwidth]{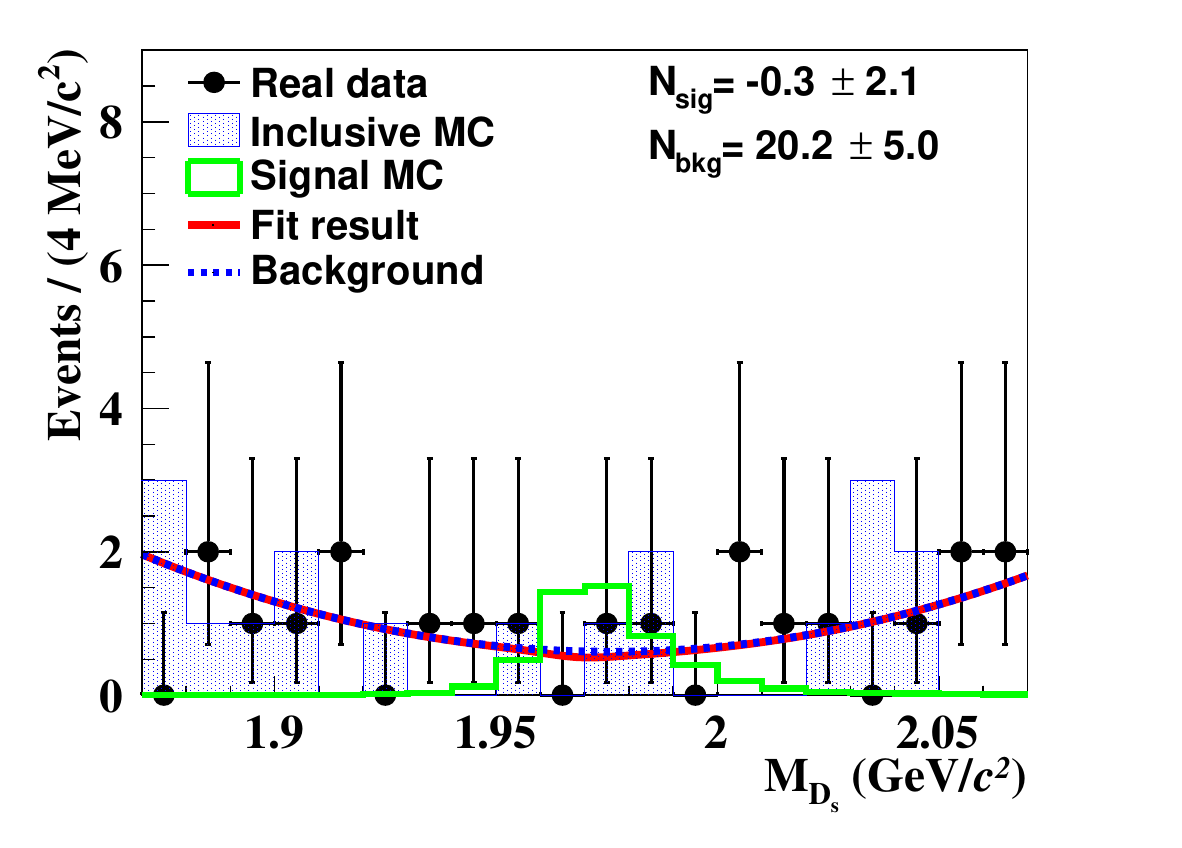}}
\caption{The $M_{D_{s}}$ distributions of the (a) $J/\psi\to D_{s}^{-}\rho^{+}$ and (b) $J/\psi\to D_{s}^{-}\pi^{+}$ candidate events. The black dots with error bars are data; the blue histograms are the inclusive MC samples; the green solid lines are the signal MC samples, scaled to the obtained 90\% UL BFs. The red line is the total fit result with the signal shape and the blue background shape dotted line.}
\label{fig:Dsrhopi}
\end{figure*}

Full reconstruction of $D_{s}$ mesons with non-leptonic-decay modes does not offer good sensitivity due to the large hadronic background from $J/\psi$ inclusive decays. Therefore, $D_{s}$ candidates are reconstructed via the semi-leptonic decay $D_{s}^{-}\to \phi e^{-} \Bar{\nu}_{e}$ with $\phi\to K^{+}K^{-}$. For the $\rho^{+}$ side, the subsequent signal decays are $\rho^{+}\to \pi^{+}\pi^{0}$, $\pi^{0}\to\gamma\gamma$. Unbinned extended maximum likelihood fits are performed to the distributions of the mass recoiling against the $\pi^{+}$ or $\rho^{+}$, denoted as $M_{D_{s}}$, to determine the signal yields, as shown in Fig.~\ref{fig:Dsrhopi}. No significant signal is observed and ULs on their BFs are set at $\mathcal{B}(J/\psi\to D_{s}^{-}\rho^{+})<8.0\times10^{-7}$ and $\mathcal{B}(J/\psi\to D_{s}^{-}\pi^{+})<4.1\times10^{-7}$ at the 90\% C.L. In comparison to the previous best limits, the UL for $J/\psi\to D_{s}^{-}\rho^{+}$ has been improved by about an order of magnitude, and the UL for $J/\psi\to D_{s}^{-}\pi^{+}$ has been improved by about three orders of magnitude.

\section{Summary}
\label{sec:summary}
Using about 10 billion $J/\psi$ events and $7.33 fb^{-1}$ $D_{s}D_{s}^{*}$ events collected by BESIII, we search for the FCNC processes $J/\psi\to D^{0}\mu^{+}\mu^{-}$, $J/\psi\to D^{0}\gamma$, and $D_{s}^{+}\to h(h')e^{+}e^{-}$. We also search for $J/\psi$ weak decays containing a D meson, $J/\psi\to D_{s}^{-}\pi^{+}$, and $J/\psi\to D_{s}^{-}\rho^{+}$. The $D_{s}^{+}\to\pi^{+}\phi(\to e^{+}e^{-})$ and $D_{s}^{+}\to\rho^{+}\phi(\to e^{+}e^{-})$ decays are observed and the BFs of these two decays are measured to be $\mathcal{B}(D_{s}^{+}\to\pi^{+}\phi, \phi\to e^{+}e^{-})=(1.17_{-0.21}^{+0.23}\pm0.03)\times10^{-5}$ and $\mathcal{B}(D_{s}^{+}\to\rho^{+}\phi, \phi\to e^{+}e^{-})=(2.44_{-0.62}^{+0.67}\pm0.16)\times10^{-5}$. No significant signal is observed for other decays. The upper limits on the BFs of these decays are set to be $\mathcal{B}(J/\psi\to D^{0}\mu^{+}\mu^{-})= 1.1 \times 10^{-7}$, $\mathcal{B}(J/\psi\to D^{0}\gamma) = 9.1 \times 10^{-8}$, $\mathcal{B}(D_{s}^{+}\to \pi^{+}\pi^{0}e^{+}e^{-})<7.0\times10^{-5}$, $\mathcal{B}(D_{s}^{+}\to K^{+}\pi^{0}e^{+}e^{-})<7.1\times10^{-5}$, $\mathcal{B}(D_{s}^{+}\to K_{S}^{0}\pi^{+}e^{+}e^{-})<8.1\times10^{-5}$, $\mathcal{B}(J/\psi\to \bar{D}^{0}\pi^{0}+c.c.)<4.7\times10^{-7}$, $\mathcal{B}(J/\psi\to \bar{D}^{0}\eta+c.c.)<6.8\times10^{-7}$, $\mathcal{B}(J/\psi\to \bar{D}^{0}\rho^{0}+c.c.)<5.2\times10^{-7}$, $\mathcal{B}(J/\psi\to D^{-}\pi^{+}+c.c.)<7.0\times10^{-8}$, $\mathcal{B}(J/\psi\to D^{-}\rho^{+}+c.c.)<6.0\times10^{-7}$, $\mathcal{B}(J/\psi\to D_{s}^{-}\rho^{+})<8.0\times10^{-7}$, and $\mathcal{B}(J/\psi\to D_{s}^{-}\pi^{+})<4.1\times10^{-7}$ at the 90\% confidence level. These results are consistent with the SM and provide constraints on some new physical models.

\section*{Acknowledgements}
This work is supported by the National Key R\&D Program of China under Contracts Nos. 2023YFA1606000; National Natural Science Foundation of China~(Grant No. 12175321, U1932101, 11975021, 11675275).

\end{document}